\documentclass[conference]{IEEEtran}
\usepackage[utf8]{inputenc}
\usepackage{amsmath}
\usepackage{graphicx}
\usepackage{algorithm}
\usepackage{algpseudocode}
\usepackage{hyperref} 
\usepackage{cite}
\usepackage{multirow}



\usepackage{eso-pic}
\newcommand\AtPageUpperMyright[1]{\AtPageUpperLeft{
 \put(\LenToUnit{0.5\paperwidth},\LenToUnit{-1cm}){
     \parbox{0.5\textwidth}{\raggedleft\fontsize{9}{11}\selectfont #1}}
 }}
\newcommand{\conf}[1]{
\AddToShipoutPictureBG*{
\AtPageUpperMyright{#1}
}
}

\begin{document}

\title{PhishGuard: A Multi-Layered Ensemble Model for Optimal Phishing Website Detection}

\conf{This work has been submitted to the IEEE for possible publication. Copyright may be transferred without notice, after which this version may no longer be accessible.} 


\author{\IEEEauthorblockN{Md Sultanul Islam Ovi\IEEEauthorrefmark{1}, Md. Hasibur Rahman\IEEEauthorrefmark{2}, and Mohammad Arif Hossain\IEEEauthorrefmark{2}}
\IEEEauthorblockA{\IEEEauthorrefmark{1}Dept. of Computer Science, George Mason University, Fairfax, Virginia, USA}
\IEEEauthorblockA{\IEEEauthorrefmark{2}Dept. of Computer Science and Engineering, Green University of Bangladesh, Bangladesh}
\IEEEauthorblockA{Email: movi@gmu.edu, 212902018@student.green.ac.bd, arifhossain.cse212@gmail.com}
}

\maketitle
\begin{abstract}
Phishing attacks are a growing cybersecurity threat, leveraging deceptive techniques to steal sensitive information through malicious websites. To combat these attacks, this paper introduces \textit{PhishGuard}, an optimal custom ensemble model designed to improve phishing site detection. The model combines multiple machine learning classifiers, including Random Forest, Gradient Boosting, CatBoost, and XGBoost, to enhance detection accuracy. Through advanced feature selection methods such as \textit{SelectKBest} and \textit{RFECV}, and optimizations like hyperparameter tuning and data balancing, the model was trained and evaluated on four publicly available datasets. \textit{PhishGuard} outperformed state-of-the-art models, achieving a detection accuracy of 99.05\% on one of the datasets, with similarly high results across other datasets. This research demonstrates that optimization methods in conjunction with ensemble learning greatly improve phishing detection performance.
\end{abstract}


\section{Introduction}

Phishing is one of the most persistent and dangerous cybercrimes, in which perpetrators trick victims into disclosing private information including bank account information, login passwords, and personal information. These attacks typically involve creating counterfeit websites that mimic legitimate ones, tricking users into trusting them. Over the years, phishing techniques have evolved, becoming increasingly sophisticated and posing a serious threat to internet users worldwide \cite{al2021optimized}. With the digital transformation and the rise in online services, phishing attacks have surged, targeting both individual users and organizations \cite{abdul2023analysis}.

Traditional phishing detection methods, such as blacklist-based approaches, are often outdated and struggle to identify zero-day phishing attacks. Heuristic methods, while more dynamic, can lack real-time adaptability \cite{atlam2022business}. These limitations have redirected attention toward machine learning (ML), which utilizes data-driven models to classify phishing websites by analyzing patterns in website features \cite{salihovic2019role}.

By using large datasets of authentic and fraudulent websites, machine learning algorithms have shown great promise in the identification of phishing attempts. Classifiers such as Random Forest (RF), Support Vector Machines (SVM), and Artificial Neural Networks (ANN) have achieved high accuracy in detecting phishing websites \cite{khan2020phishing}. Moreover, ensemble learning methods, which combine multiple classifiers, have gained traction in improving detection accuracy and generalization. 

In this paper, we introduce \textbf{PhishGuard}, a multi-layered custom ensemble model designed for optimal phishing website detection. PhishGuard integrates several machine learning models in a stacked ensemble architecture. Additionally, this research emphasizes the importance of feature selection techniques such as Recursive Feature Elimination with Cross-Validation (RFECV) and Principal Component Analysis (PCA) \cite{jolliffe2016principal}, optimizing the performance of phishing detection models across multiple datasets, including the UCI\cite{mohammad2012assessment} and Mendeley phishing datasets \cite{tan2018phishing, hannousse2021web}.

This paper presents a robust ensemble approach that surpasses existing methods, demonstrating the effectiveness of optimized machine learning in phishing detection across diverse datasets.

\section{Literature Review}

The shortcomings of conventional phishing detection systems, such as those that rely on heuristics and blacklists, have been effectively addressed by machine learning (ML) \cite{atlam2022business}. Traditional methods have demonstrated limited effectiveness in identifying novel or sophisticated phishing attacks, particularly in real-time protection. This gap has driven the adoption of ML-based methods that focus on data-driven classification of phishing webpages.

\subsection{Machine Learning Techniques in Phishing Detection}

A thorough comparison of several machine learning techniques, such as Decision Trees (DT), Support Vector Machines (SVM), Random Forest (RF), Naïve Bayes (NB), k-Nearest Neighbors (kNN), and Artificial Neural Networks (ANN), was carried out on three different datasets, including the UCI Phishing Dataset, by Khan et al. \cite{khan2020phishing}. According to their results, ANN and Random Forest were two of the best models. Because models with optimized feature sets outperformed those without, the study also demonstrated the significance of careful feature selection. Similar research was conducted by Salihovic et al. \cite{salihovic2019role}, who evaluated several algorithms on a variety of datasets and discovered that feature selection techniques, such as Principal Component Optimization, greatly enhanced the identification of phishing websites.

Feature selection has been widely recognized as an essential component in improving the performance of phishing detection models. Hutchinson et al. \cite{hutchinson2018detecting} explored feature grouping, categorizing features into URL-based, host-based, and ranking-based categories. Their experiments revealed that feature sets combining multiple categories led to improved performance. Additionally, Vishva et al. \cite{vishva2022phisher} utilized the TF-IDF NLP technique to combine URL and webpage content features, with promising results when applied to phishing detection.

Recent studies have further explored the role of fine-tuning machine learning models to maximize their performance. For instance, Abdul Samad et al. \cite{abdul2023analysis} demonstrated that fine-tuning techniques, such as hyperparameter optimization and feature selection, can significantly enhance the effectiveness of machine learning models. Their research on phishing URL detection found that these optimizations led to a marked improvement in detection performance, further highlighting the importance of selecting the right features and optimizing model parameters.

\subsection{Ensemble Learning in Phishing Detection}

Ensemble learning methods, which combine multiple classifiers to achieve better generalization, have been increasingly adopted for phishing detection. Sarasjati et al. \cite{sarasjati2022comparative} compared multiple ensemble methods, such as XGBoost, Gradient Boosting Machines (GBM), and Rotation Forest, with Random Forest as a baseline. Their findings support the argument that ensemble models, particularly those incorporating decision trees, can offer superior performance in phishing detection tasks.

Al-Sarem et al. \cite{al2021optimized} introduced a stacking ensemble method that used multiple classifiers for phishing website detection. Their approach integrated classifiers such as Random Forest, SVM, and kNN, showing the potential to enhance performance by harnessing the strengths of multiple classifiers and minimizing the chances of misclassifying phishing websites.

\subsection{Research Gaps and Future Directions}

While the literature underscores the effectiveness of machine learning models for phishing detection, there remain several areas requiring further research. A key challenge is improving feature selection techniques to enhance model performance while minimizing computational complexity. Both Al-Sarem et al. \cite{al2021optimized} and Sarasjati et al. \cite{sarasjati2022comparative} emphasized the need for optimized feature selection, particularly in reducing false positives.

Moreover, while ensemble learning methods like stacking and boosting have shown considerable promise, further research is required to assess their effectiveness across diverse datasets, including newer, more complex phishing datasets. Abdul Samad et al. \cite{abdul2023analysis} also highlighted the need for models to be fine-tuned across various datasets to ensure generalization and robustness. Models need to be developed and adapted to handle a wide variety of datasets with differing characteristics, as current models often face difficulties generalizing beyond the datasets they were trained on.

\begin{figure}[htbp]
\centering
\includegraphics[width=\linewidth]{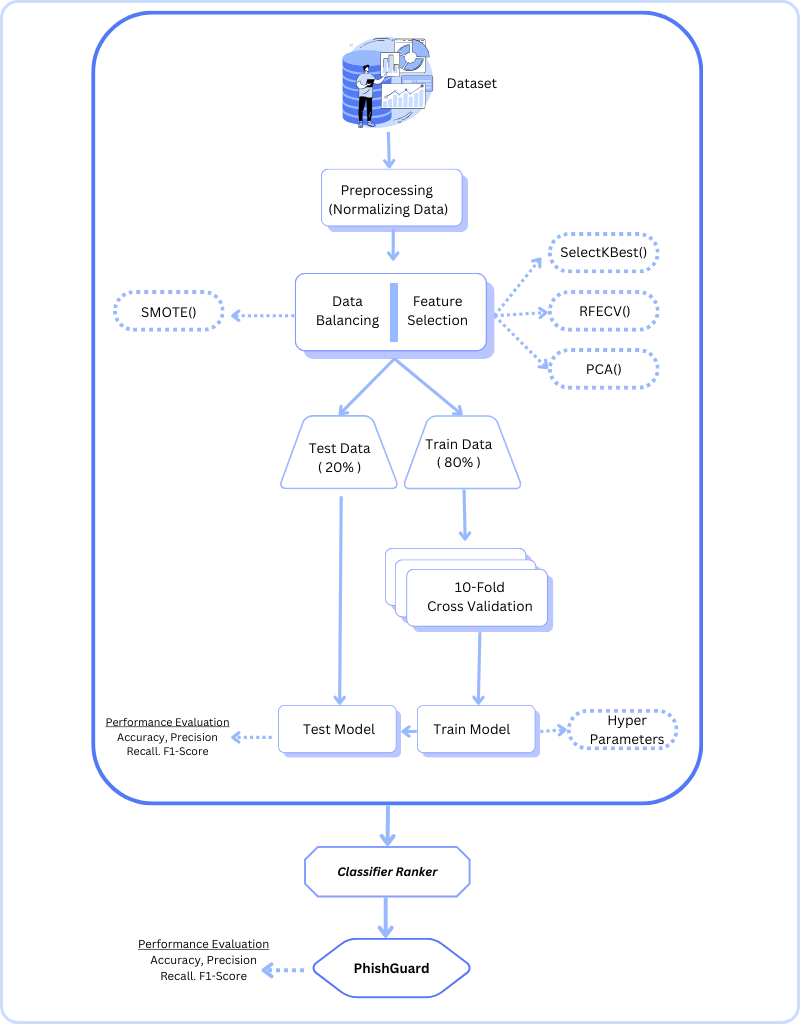}
\caption{Overview of the PhishGuard development process}
\label{fig:process_overview}
\end{figure}

\section{Methodology}
The PhishGuard model is a multi-layered ensemble classifier that is intended for the best possible identification of phishing websites. In this part, we outline the procedure used to build it. An overview of the complete procedure used in this investigation is shown in Figure \ref{fig:process_overview}. The four primary stages of the methodology are feature selection, data preprocessing, model training, and building the customized ensemble model.

During the data preprocessing phase, normalization and SMOTE techniques \cite{chawla2002smote} were applied to balance the datasets and ensure that the models were trained on unbiased data. In the feature selection phase, three methods—SelectBest, RFECV, and PCA—were used to retain the most important features for phishing detection. Following this, six machine learning models were trained and optimized through hyperparameter tuning. These models were then ranked based on their evaluation metrics, and the top-performing models were selected to construct the PhishGuard stacked ensemble.

The final phase involved training PhishGuard, where the first-ranked model was used as the meta-model, and the 2nd, 3rd, and 4th-ranked models served as base models. By using this method, we were able to enhance the detection performance overall by combining the advantages of the separate models. To confirm PhishGuard's efficacy, it was tested against both standalone models and other relevant works after training. The subsequent sections provide a detailed explanation of each phase.


\begin{table}[htbp]
\centering
\caption{Overview of Datasets used in the Study}
\label{tab:datasets}
\begin{tabular}{|l|c|c|}
\hline
\textbf{Name of the Dataset} & \textbf{\# Instances} & \textbf{\# Features} \\ \hline
Phishing Dataset for ML\cite{tan2018phishing} & 10,000 & 48 \\ \hline
Phishing Websites (UCI)\cite{mohammad2012assessment} & 11,055 & 30 \\ \hline
Web Page Phishing Detection\cite{hannousse2021web} & 11,430 & 87 \\ \hline
PhishStorm\cite{marchal2014phishstorm} & 96,018 & 12 \\ \hline
\end{tabular}
\end{table}

\subsection{Data Collection and Dataset Description}

For this study, we utilized four publicly available phishing detection datasets, each offering a unique set of features and balanced distributions between phishing and legitimate URLs. These datasets provide a comprehensive view of phishing detection mechanisms and contribute to the robustness of our proposed PhishGuard model. Table \ref{tab:datasets} provides an overview of the key characteristics of each dataset used in this study. Below is a detailed description of each dataset used in the study:

Choon Lin Tan assembled the first dataset, \textit{Phishing Dataset for Machine Learning: Feature Evaluation [Dataset 1]}\cite{tan2018phishing}, which comprises 10,000 URLs that are evenly divided between phishing and trustworthy websites. This dataset includes 48 features extracted using Selenium WebDriver, a more precise and robust feature extraction technique than conventional parsing methods.

The second dataset, called \textit{Phishing Websites Dataset [Dataset 2]}, has 11,055 cases with 30 attributes that were contributed by researchers to the UCI repository \cite{mohammad2012assessment}. There are 6,157 authentic websites and 4,898 fraudulent websites in the sample.

Published by Hannousse and Yahiouche \cite{hannousse2021web}, the third dataset, \textit{Web Page Phishing Detection [Dataset 3]}, comprises 11,430 URLs with 87 attributes extracted from them. The dataset is entirely balanced, with 50\% of the URLs being phished and 50\% being genuine. For phishing detection systems that rely on machine learning, this dataset acts as a standard.

Samuel Marchal et al. \cite{marchal2014phishstorm} contributed the final dataset, \textit{PhishStorm [Dataset 4]}, which consists of 96,018 URLs split equally between 48,009 phishing and 48,009 genuine URLs.

\begin{algorithm}[htbp]
\caption{Preprocessing for Machine Learning Models}
\label{alg:preprocessing}
\begin{algorithmic}[1]
\State \textbf{Input:} Dataset $D = (X, y)$
\State \textbf{Output:} Preprocessed data
\State \textbf{Step 1: Data Partitioning}
    \State Split data: $X_{train}, X_{test}, y_{train}, y_{test}$
    
\State \textbf{Step 2: Normalization}
    \State $X_{train\_norm} \gets \text{scaler.fit\_transform}(X_{train})$
    \State $X_{test\_norm} \gets \text{scaler.transform}(X_{test})$
    
\State \textbf{Step 3: SMOTE Balancing}
    \State $X_{train\_smote}, y_{train\_smote} \gets \text{smote.fit\_resample}(X_{train}, y_{train})$

\State \textbf{Step 4: Feature Selection}
    \State \textbf{4.1: SelectKBest}
    \For{$k = 1$ to num\_features}
        \State Apply $k$-best: $X_{train\_kbest} \gets kbest.fit\_transform(X_{train\_smote}, y_{train\_smote})$
    \EndFor

    \State \textbf{4.2: RFECV}
    \State $X_{train\_rfe}, X_{test\_rfe} \gets rfecv.fit\_transform(X_{train\_kbest}, y_{train\_smote})$

\State \textbf{Step 5: Dimensionality Reduction}
    \State $X_{train\_pca} \gets \text{pca.fit\_transform}(X_{train\_rfe})$
    \State $X_{test\_pca} \gets \text{pca.transform}(X_{test\_rfe})$
    
\end{algorithmic}
\end{algorithm}

\subsection{Data Balancing}
Phishing website detection frequently encounters imbalanced datasets, where the majority class (legitimate websites) dominates, resulting in biased models that underperform on the minority class (phishing websites) \cite{ul2022significance, zheng2022method}. The Synthetic Minority Oversampling Technique (SMOTE)\cite{chawla2002smote} was used during preprocessing in order to solve this problem. SMOTE generates synthetic samples for the minority class, balancing the dataset without overfitting, unlike random oversampling. The data balancing step, including SMOTE, was implemented as outlined in the algorithm. This approach was crucial in mitigating the effects of dataset imbalance, allowing the model to perform effectively across both classes. This led to improved phishing webpage identification, which is the major objective of this study, while classification performance remained impartial toward genuine sites.


\subsection{Feature Selection and Dimensionality Reduction}
To enhance model performance, a two-step feature selection process was applied, followed by PCA for dimensionality reduction.

\textbf{Feature Selection}

Initially, the \textit{SelectKBest} method was utilized, using ANOVA F-value to identify the most significant features \cite{guyon2003introduction}. To find the ideal number of features, a grid search was conducted and cross-validation was used for every model (e.g., Random Forest, SVM). The best set of features was then selected for further training and testing.

Then, to further hone the feature set and choose the most crucial characteristics for every model, Recursive Feature Elimination with Cross-Validation (RFECV)\cite{guyon2002gene} was used. This ensured that the models focused on the most relevant features, improving accuracy.

\begin{table*}[htbp]
\caption{Evaluation Metrics(\%) Across Four Datasets}
\centering
\begin{tabular}{|l|cccc|cccc|cccc|cccc|}
\hline
\multirow{2}{*}{\textbf{Model}} & \multicolumn{4}{|c|}{\textbf{Dataset 1}} & \multicolumn{4}{|c|}{\textbf{Dataset 2}} & \multicolumn{4}{|c|}{\textbf{Dataset 3}} & \multicolumn{4}{|c|}{\textbf{Dataset 4}} \\
\cline{2-17}
 & \textbf{Acc.} & \textbf{Prec.} & \textbf{Rec.} & \textbf{F1} & \textbf{Acc.} & \textbf{Prec.} & \textbf{Rec.} & \textbf{F1} & \textbf{Acc.} & \textbf{Prec.} & \textbf{Rec.} & \textbf{F1} & \textbf{Acc.} & \textbf{Prec.} & \textbf{Rec.} & \textbf{F1} \\
\hline
SVM & 97.30 & 96.87 & 97.83 & 97.35 & 96.61 & 96.24 & 97.85 & 97.04 & 96.19 & 95.81 & 96.72 & 96.26 & 81.65 & 80.45 & 83.68 & 82.04 \\
RF & 98.60 & 98.52 & 98.72 & 98.62 & 97.02 & 96.63 & 98.17 & 97.39 & 97.11 & 96.58 & \textbf{97.75} & 97.17 & \textbf{95.13} & \textbf{95.69} & 94.53 & \textbf{95.11} \\
\hline
XGB & \textbf{99.00} & \textbf{98.82} & \textbf{99.21} & \textbf{99.01} & 96.70 & 96.61 & 97.61 & 97.11 & 97.11 & 96.99 & 97.32 & 97.15 & 94.97 & 95.00 & \textbf{94.95} & 94.98 \\
CB & 98.65 & 98.24 & 99.11 & 98.67 & \textbf{97.24} & \textbf{96.64} & 98.57 & \textbf{97.59} & 97.20 & 96.91 & 97.58 & 97.24 & 94.06 & 94.28 & 93.84 & 94.06 \\
\hline
AB & 97.50 & 97.16 & 97.93 & 97.54 & 93.80 & 94.51 & 94.58 & 94.54 & 95.93 & 95.86 & 96.11 & 95.99 & 87.78 & 87.17 & 88.64 & 87.90 \\
GB & 98.85 & 98.53 & \textbf{99.21} & 98.87 & 97.11 & 96.34 & \textbf{98.65} & 97.48 & \textbf{97.29} & \textbf{97.08} & 97.58 & \textbf{97.33} & 94.79 & 94.77 & 94.84 & 94.80 \\
\hline
\end{tabular}
\label{tab:combined_metrics}
\end{table*}

\textbf{Dimensionality Reduction}

Following feature selection, PCA\cite{jolliffe2016principal} was used to minimize the feature space's dimensionality while preserving at least 95\% of the data's variance. This step reduced model complexity while preserving critical information.

The combination of \textit{SelectKBest}, \textit{RFECV}, and \textit{PCA} allowed the models to be trained on an optimal, low-dimensional feature set, enhancing phishing detection performance \cite{li2017feature}.

\subsection{Machine Learning Algorithms}
In the PhishGuard model, a variety of machine learning algorithms were employed to classify phishing websites. Below is a brief overview of the models used:

\textbf{Support Vector Machine (SVM)} \cite{boser1992training} is effective in high-dimensional spaces, using different kernel functions to find the optimal hyperplane separating phishing from legitimate sites.

\textbf{Random Forest (RF)} \cite{breiman2001random} builds multiple decision trees, aggregating predictions for strong generalization. Its robustness against overfitting makes it reliable for phishing detection.

\textbf{XGBoost (XGB)} \cite{chen2016xgboost} is a gradient boosting algorithm known for speed and accuracy. Its ability to handle sparse data and regularization makes it ideal for phishing classification.

\textbf{CatBoost (CB)} \cite{prokhorenkova2018catboost} handles categorical features efficiently, excelling in imbalanced datasets with minimal tuning.

\textbf{AdaBoost (AB)} \cite{freund1997decision} combines weak classifiers into a strong model by iteratively focusing on misclassified instances.

\textbf{Gradient Boosting (GB)} \cite{friedman2001greedy} sequentially builds decision trees, optimizing the loss function to fit complex patterns, improving phishing detection accuracy.

These algorithms were trained and evaluated, contributing to the final ensemble model for phishing detection.

\subsection{Design of the Customized Ensemble Model}
The PhishGuard model employs a customized ensemble learning approach, combining multiple machine learning models to enhance performance. After ranking the optimized classifiers, the 2nd, 3rd, and 4th best-performing models were selected as base classifiers, while the top-ranked model was used as the meta-model in a stacking method \cite{wolpert1992stacked}.

In this stacking approach, the base models' predictions on the training data were used as inputs for the meta-model, leveraging the strengths of each model to improve accuracy and robustness in phishing detection. Stacking allows for the combination of diverse models, reducing prediction errors and enhancing overall performance across various test cases \cite{rokach2010ensemble}.

\subsection{Hyperparameter Tuning}
Hyperparameter tuning was essential for optimizing the six models used in PhishGuard. For \textbf{SVM}, the regularization parameter ($C$), kernel type, and margin width were adjusted. In \textbf{Random Forest}, the number of trees, tree depth, and features per split were fine-tuned. For \textbf{XGBoost} and \textbf{CatBoost}, tuning focused on learning rate, tree depth, and iterations, while \textbf{AdaBoost} was optimized by adjusting the number of estimators and learning rate. \textbf{Gradient Boosting} fine-tuned learning rate, estimators, and tree depth.

Grid search and random search were used to explore the best hyperparameter combinations, with cross-validation ensuring optimal performance on unseen phishing detection tasks \cite{kohavi1995study}.

\subsection{Model Evaluation Metrics}
Four essential measures were used to assess PhishGuard's performance: F1-Score, Accuracy, Precision, and Recall. The overall accuracy of the model's predictions is indicated by \textbf{Accuracy}. Out of all predicted phishing occurrences, \textbf{Precision} shows the percentage of accurately identified phishing pages, and \textbf{Recall} gauges how well the model can identify real phishing sites. \textbf{F1-Score} provides a useful statistic for imbalanced data by striking a balance between precision and recall.

These metrics assess how effectively the model generalizes to yet-undiscovered data and how accurate it is in identifying fraudulent websites. Given the evolving nature of phishing efforts, these evaluation criteria are crucial to verifying that PhishGuard can maintain dependable detection throughout a broad spectrum of dynamic phishing scenarios.
\section{Results and Discussion}

Now that we have carefully adhered to the research methods described before in this work, we present and discuss the findings of our experiments.

\subsection{Performance of Machine Learning Classifiers}

The results of the ML classifiers on four datasets are compiled in Table \ref{tab:combined_metrics}. XGB consistently achieved the highest accuracy on Dataset 1 (99.00\%) and performed well across all datasets. RF excelled in Dataset 3 with the highest recall (97.75\%) and demonstrated robustness across other datasets. CB showed strong performance, particularly in Dataset 2 with the highest accuracy (97.24\%) and recall (98.57\%). Although AB’s performance was slightly lower, it was still competitive on Dataset 3 (95.93\% accuracy). SVM maintained consistent results but struggled on Dataset 4, where its accuracy dropped to 81.65\%. GB delivered reliable results, notably the best F1 score on Dataset 3 (97.33\%). Based on these results, XGB, RF, CB, and GB were selected for the ensemble model, as they consistently performed well across different datasets.

\subsection{Effectiveness of the Customized Ensemble Model}

Table \ref{tab:accuracy_datasets} presents the accuracies of individual models across datasets. Our customized ensemble model, PhishGuard, consistently outperformed other models, achieving the highest accuracy on all datasets. For each dataset, the best-performing model was selected as the meta-learner, and the second, third, and fourth models as base learners. For example, XGB was the meta-learner for Dataset 01, while CB was chosen for Dataset 02. This strategic selection boosted PhishGuard’s performance, with 99.05\% accuracy on Dataset 01 and 95.17\% on Dataset 04, showcasing its effectiveness in phishing detection.

\begin{table}[htbp]
\caption{Accuracies(\%) of Machine Learning Models Across Different Datasets}
\centering
\begin{tabular}{|l|c|c|c|c|}
\hline
\textbf{ML Model} & \textbf{Dataset 01} & \textbf{Dataset 02} & \textbf{Dataset 03} & \textbf{Dataset 04} \\
\hline
SVM & 97.30 & 96.61 & 96.19 & 81.65 \\
RF & 98.60 & 97.02 & 97.11 & 95.13 \\
\hline
XGB & 99.00 & 96.70 & 97.11 & 94.97 \\
CB & 98.65 & 97.24 & 97.20 & 94.06 \\
\hline
AB & 97.50 & 93.80 & 95.93 & 87.78 \\
GB & 98.85 & 97.11 & 97.29 & 94.79 \\
\hline
\textbf{PhishGuard} & \textbf{99.05} & \textbf{97.29} & \textbf{97.33} & \textbf{95.17} \\
\hline
\end{tabular}
\label{tab:accuracy_datasets}
\end{table}
\begin{figure}[htbp]
\centering
\includegraphics[width=\linewidth]{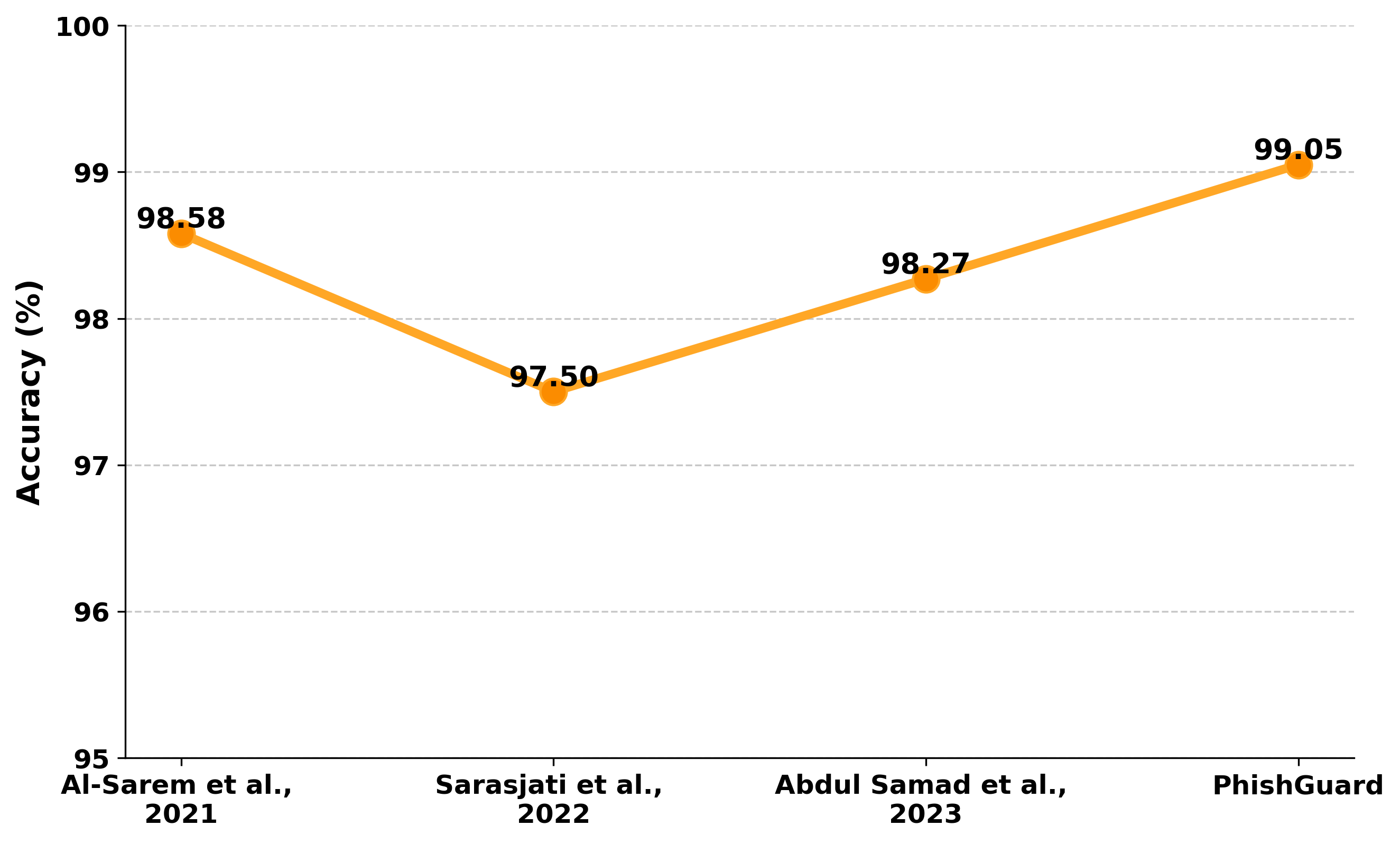}
\caption{Comparative analysis of PhishGuard against advanced algorithms \cite{al2021optimized,sarasjati2022comparative,abdul2023analysis} demonstrates that PhishGuard surpasses all existing approaches on Dataset 01.}
\label{fig:comparative_analysis}
\end{figure}
\begin{figure}[htbp]
\centering
\includegraphics[width=\linewidth]{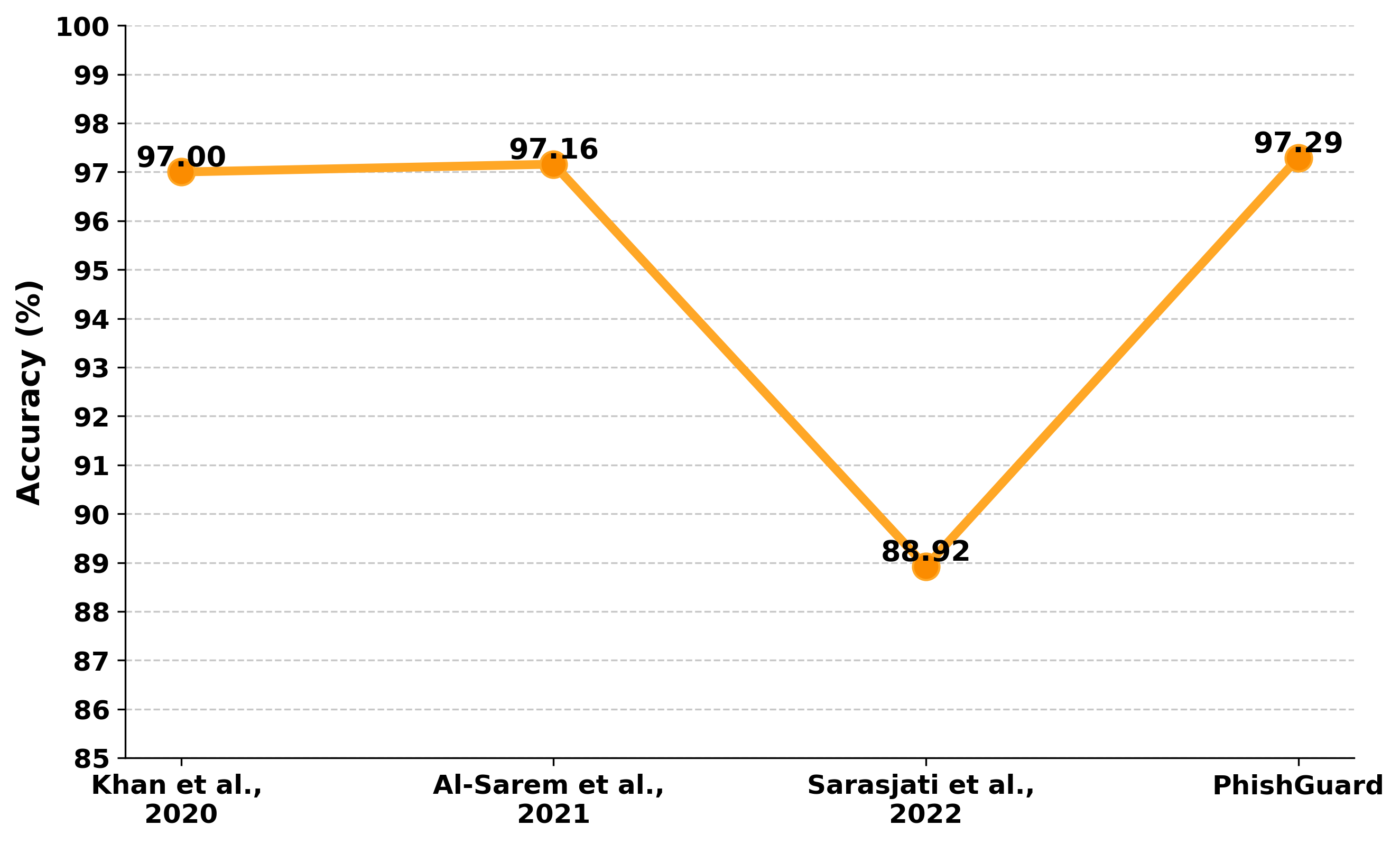}
\caption{Comparative analysis of PhishGuard against advanced algorithms \cite{khan2020phishing,al2021optimized,sarasjati2022comparative} demonstrates that PhishGuard surpasses all existing approaches on Dataset 02.}
\label{fig:comparative_analysis_dataset2}
\end{figure}

\subsection{Comparison Study Using Cutting-Edge Models}

On every dataset, PhishGuard regularly outperforms the most recent models. For \textit{Dataset 1}, PhishGuard achieved the highest accuracy of 99.05\%, surpassing models such as Al-Sarem et al. \cite{al2021optimized} (98.58\%), Sarasjati et al. \cite{sarasjati2022comparative} (97.50\%), and Abdul Samad et al. \cite{abdul2023analysis} (98.27\%) as shown in Figure 2. Similarly, for \textit{Dataset 2}, PhishGuard recorded 97.29\%, exceeding Khan \cite{khan2020phishing} (97.00\%), Al-Sarem \cite{al2021optimized} (97.16\%), and Sarasjati \cite{sarasjati2022comparative} (88.92\%) as displayed in Figure 3. While fewer studies exist for \textit{Dataset 3} and \textit{Dataset 4}, PhishGuard still outperformed the available models. It achieved 97.33\% on \textit{Dataset 3}, surpassing Uddin et al. \cite{uddin2022comparative} (97.00\%), and 95.17\% on \textit{Dataset 4}, exceeding Marchal et al. \cite{marchal2014phishstorm} (94.91\%). In summary, PhishGuard consistently demonstrates superior performance, even in under-researched datasets.

\section{Conclusion}
This study introduced \textit{PhishGuard}, a customized ensemble model for phishing website detection. By combining Gradient Boosting, Random Forest, and XGBoost, the model leveraged the strengths of each classifier and utilized feature selection techniques like \textit{SelectKBest} and \textit{RFECV} to enhance its detection capabilities. The results across multiple datasets demonstrate that \textit{PhishGuard} consistently outperforms traditional machine learning methods, providing superior phishing detection performance. Future research can focus on integrating real-time phishing data for dynamic model training, exploring advanced ensemble techniques, and adapting \textit{PhishGuard} for use in \textit{IoT} environments. These efforts will further improve the model’s adaptability and resilience against evolving phishing threats, ensuring that it remains a robust tool in cybersecurity.


\end{document}